\pdfoutput=1

\documentclass[11pt]{article}

\usepackage[final]{acl}

\usepackage{times}
\usepackage{latexsym}

\usepackage[T1]{fontenc}

\usepackage[utf8]{inputenc}

\usepackage{microtype}

\usepackage{inconsolata}

\usepackage{graphicx}

\usepackage{enumitem}
\usepackage{amsmath} 
\usepackage{tcolorbox}
\usepackage{graphicx}
\usepackage{amsfonts}
\usepackage{graphicx}
\usepackage{subcaption}

\title{Amplifying Your Social Media Presence: \\
Personalized Influential Content Generation with LLMs}

\author{
 \textbf{Yuying Zhao\textsuperscript{1}},
 \textbf{Yu Wang\textsuperscript{2}},
 \textbf{Xueqi Cheng\textsuperscript{1}},
 \textbf{Anne Marie Tumlin\textsuperscript{1}},
\\
 \textbf{Yunchao Liu\textsuperscript{1}},
 \textbf{Damin Xia\textsuperscript{1}},
 \textbf{Meng Jiang\textsuperscript{3}},
 \textbf{Tyler Derr \textsuperscript{1}}
\\
\\
 \textsuperscript{1}Vanderbilt University,
 \textsuperscript{2}University of Oregon,
 \textsuperscript{3}University of Notre Dame
\\
 \small{
   \texttt{\{yuying.zhao, xueqi.cheng, anne.m.tumlin, yunchao.liu, damin.xia, tyler.derr\}@vanderbilt.edu}
 }
 \\
 \small{\texttt{yuwang@uoregon.edu, mjiang2@nd.edu}}
}

\begin{document}
\maketitle
\begin{abstract}
The remarkable advancements in Large Language Models (LLMs) have revolutionized the content generation process in social media, offering significant convenience in writing tasks. However, existing applications, such as sentence completion and fluency enhancement, do not fully address the complex challenges in real-world social media contexts. A prevalent goal among social media users is to increase the visibility and influence of their posts. This paper, therefore, delves into the compelling question: \textit{Can LLMs generate personalized influential content to amplify a user’s presence on social media?} We begin by examining prevalent techniques in content generation to assess their impact on post influence. Acknowledging the critical impact of underlying network structures in social media, which are instrumental in initiating content cascades and highly related to the influence/popularity of a post, we then inject network information into prompt for content generation to boost the post's influence. We design multiple content-centric and structure-aware prompts. The empirical experiments across LLMs validate their ability in improving the influence and draw insights on which strategies are more effective. Our code is available at 
\url{https://github.com/YuyingZhao/LLM-influence-amplifier}
\end{abstract}

\section{Introduction}
\label{sec.intro}
Large Language Models (LLMs) have marked a significant milestone in content generation. Their applications span a diverse range of areas, including but not limited to question answering, translation, and summarization~\cite{hadi2023survey}. 
On social media platforms, users can leverage these tools to facilitate the content creation process.
For instance, a user might specify keywords or provide the initial part of a post, enabling the model to generate content based on this input. Alternatively, a user could submit the entire post, seeking the model's assistance in refining the content for enhanced coherence and fluency. These completion or refinement methods substantially reduce the writing burden of content creators, allowing users to focus more on the creative aspects rather than the writing details, thereby offering considerable convenience~\cite{radensky2024let}.

\begin{figure}
    \centering
    \includegraphics[width=0.43\textwidth]{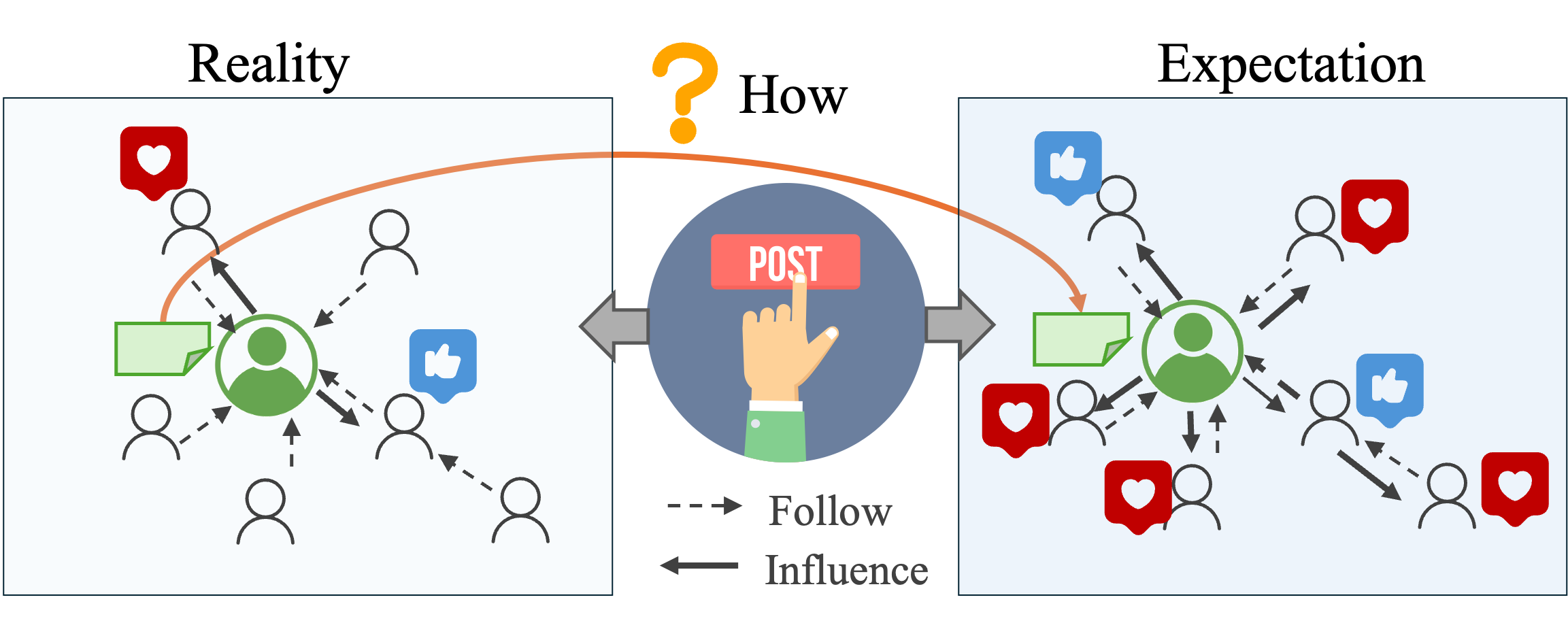}
    \vskip -1.5ex
    \caption{Many social media users strive to maximize their influence, seeking more reposts and likes. In reality this is quite challenging and they may only receive engagement from a small portion of their audience. How can a user create a post that gains larger influence?}
    \vskip -3.5ex
    \label{fig.background}
\end{figure}

Nevertheless, the objectives of completing a sentence or enhancing its fluency do not entirely encapsulate the inherent complexities encountered in real-world social media environments. In such settings, content generation is not the sole ultimate goal. Many social media users aspire to increase their posts' visibility and influence~\cite{kempe2003maximizing}. This objective differs fundamentally from generating or refining sentences which can be measured by comparing the pair of generated content and the target content. As illustrated in Fig.~\ref{fig.background}, when a user posts a piece of information, it will disseminate according to the underlying network that connect users together. This demonstrates that gaining popularity is not a simple pairwise task but a multifaceted challenge involving complex interactions among multiple users. Social media strategist can contribute to this goal of gaining popularity by analyzing the interests and behavior patterns of the target audience. However, scaling these insights to a larger audience and applying them to revise posts can be challenging in practice. In order to automate such process, we turn to LLMs.

Inspired by the success of LLMs in general content generation tasks, we explore their potential in this complex social task by posing the question:

\noindent \textit{\textbf{Can LLMs generate personalized influential content to amplify a user’s presence on social media?}}

\begin{figure}
    \centering
    \hspace*{-0.8ex}
    \includegraphics[width=0.5\textwidth]{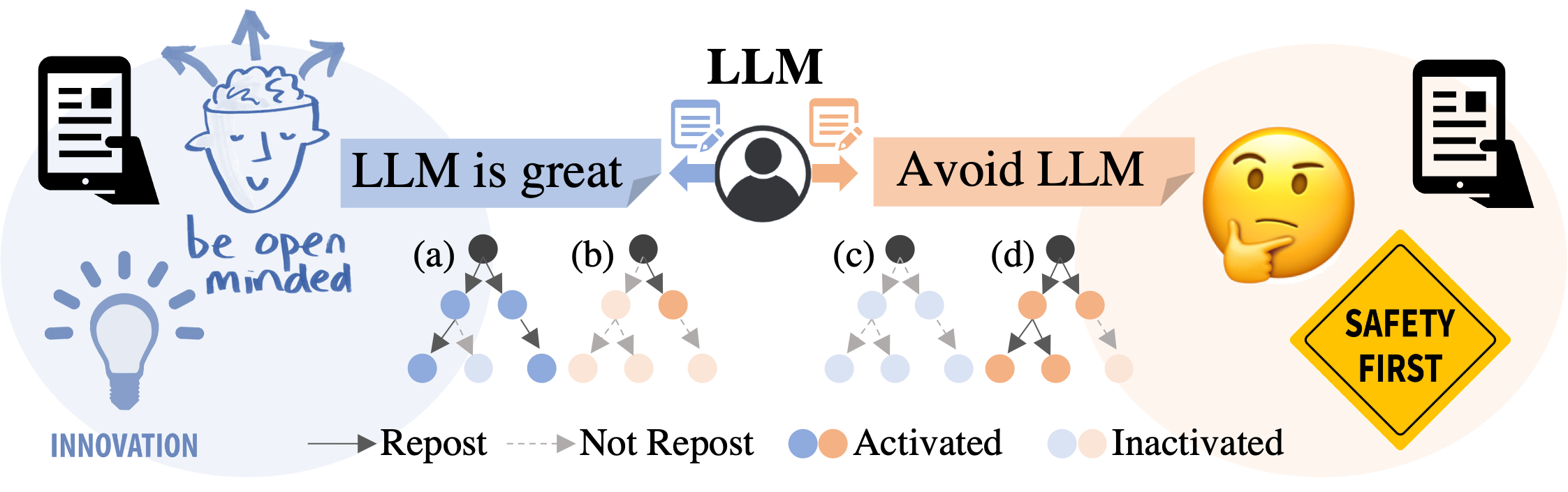}
    \caption{Online community structure matters: the same post related to LLM will spread differently in a tech-friendly versus tech-skeptical local neighborhood.}
    \label{fig:structure_matters}
    \vskip -1.5ex
\end{figure}

Addressing this question is challenging since it requires (1) a deep understanding of how generated content spreads within social networks, and (2) the development of effective strategies to amplify its dissemination. To answer this question, we first design a content-aware influence estimator which enables the influence evaluation of generated content so that we have the tool to determine which strategies are more effective. 
Equipped with the evaluator, we investigate whether the commonly utilized strategies for adapting LLMs to downstream tasks are effective in improving content influence (named as \textit{content-centric strategies}). Furthermore, the social network structure is crucial for the content propagation. In Fig.~\ref{fig:structure_matters}, we show how the online community structure (i.e., neighborhood) largely impacts the information spread through a simple example. For instance, a creator is currently thinking about writing a post about LLM. With the given keywords, the main idea of the generated content might be positive (e.g., LLM is great) or negative (e.g., Avoid LLM). The impact of these two posts vary significantly on diverse networks. If the majority of neighbors are tech-friendly, the positive post about new technology is highly likely to intrigue more reposts than a negative one in Fig.~\ref{fig:structure_matters}(a, b). A similar case happens when the majority of neighbors are tech-skeptical in Fig.~\ref{fig:structure_matters}(c, d). Inspired by the significance of structure, we design several strategies to extract local neighborhood information of the content creator and inject them into the prompting (named as \textit{structure-aware strategies}).

Our main contributions are summarized as:
\begin{itemize}[leftmargin=*]
    \vspace{-0.75ex}
    \item To the best of our knowledge, we are the initial work to explore whether LLMs help improve influence spread in social media. 
    \vspace{-0.75ex}
    \item We design a content-aware influence estimator and devise
    multiple content-centric and structure-aware prompts to amplify the social impact.
    \vspace{-0.75ex}
    \item We empirically evaluate the proposed prompts on various LLMs to validate their ability in amplifying the influence.
\end{itemize}

\section{Related Work}
\label{sec.related_work}
Here we review 
works on the estimation of influence spread and LLM applications in social media.

\subsection{Influence Estimation}
\label{sec.related_work_influence}

Influence estimation aims to approximate the number of influenced nodes, denoted as $\sigma(\mathcal{S})$, of a given seed set $\mathcal{S}$~\cite{li2018IMsurvey}. This process is initially performed based on independent cascade (IC) or linear threshold (LT) model via Monte Carlo (MC) simulation~\cite{kempe2003maximizing}, which is computationally expensive. To improve the efficiency, simulation-based speedup techniques~\cite{leskovec2007cost,goyal2011celf++}, proxy-based approach~\cite{kimura2006tractable,chen2010scalable}, and sampling-based approach~\cite{tang2014influence,nguyen2016stop,huang2017revisiting} are proposed.
In addition to these traditional methods, deep learning (DL) techniques have been developed to estimate influence on large-scale networks~\cite{li2023IMsurvey}. DeepCas~\cite{li2017deepcas} predicts the size of cascade through sampling node sequence from cascade, learning subgraph embedding with Gated Recurrent Unit (GRU) model, and predicting the size with a multilayer perceptron (MLP). DeepInf~\cite{qiu2018deepinf} predicts user-level social influence (i.e., the action status of a user given neighborhood status and local structure information) based on a convolutional network and attention network. Furthermore, there is a growing trend towards incorporating contextual information, such as topics~\cite{li2019maximizing}, locations~\cite{cai2020target}, and temporal data~\cite{meirom2021controlling}, to enhance influence estimation.
Despite these advancements, only a few study estimate influence using the detailed content which is crucial for the estimation~\cite{chen2023cminet}. CMINet~\cite{chen2023cminet} proposes a graph-based framework for content-aware multi-channel influence diffusion which is able to fully leverage content information consisting of various types (e.g., texts, images, and videos). While most traditional methods improve the efficiency of influence estimation, they are inherently non-differentiable and thus not suitable for end-to-end training. In this study, we develop context-aware method pioneered by CMINet, pretraining a model to predict the likelihood of influence while incorporating content information into training. While CMINet focuses on estimating influence given fixed content, we aim to generate content that maximizes influence.

\subsection{LLMs in Social Media}

LLMs have been extensively adopted across various downstream tasks~\cite{hadi2023survey}. Specifically, within the domain of social media, LLMs have been utilized to improve the personalized recommendation so that a higher-qualified feed will be presented to the audience~\cite{zhao2024recommender}. Additionally, LLMs have revolutionized content generation, exhibiting the capability to produce text in diverse styles~\cite{grimme2023lost, meier2024llm} and generate persuasive content~\cite{burtell2023artificial}. LLM agents have been designed to select the influencer in digital advertising campaigns with simulation of real-world dynamics~\cite{zhang2024llm}. Despite these advancements, a comprehensive analysis exploring the potential of LLMs to enhance content dissemination remains absent.

\section{Content-Aware Influence Estimator}
\label{sec.estimator_design}

To generate influential content, we need to understand the definition of influence and therefore it can be used for potential optimization and further evaluation. Equipped with the preliminary knowledge in the last section, we introduce our content-aware method to estimate the influence of a new piece of content. Specifically, we (1) train a model to estimate the pairwise influence of the probability that a user would repost a given content by another user based on the historical interactions; (2) obtain the content influence based on the commonly-used cascade model and the learned probability.

\subsection{Pairwise Influence Estimation}
The likelihood of reposting behavior is affected by post content and author identity. For example, if two creators post similar content but a viewer prefers one creator over the other, the chances of the viewer reposting the content will vary significantly. This observation leads to our design of a pairwise influence estimator $\mathcal{P}_\theta(u_r, u_c, c)$. It is trained to predict the probability that a potential recipient user $u_r$ will repost the content $c$ posted by creator $u_c$. The probability equation is:
\begin{equation*}
    \mathcal{P}_\theta(u_r, u_c, c) = \text{sigmoid}\left({\mathbf{x}_{u_r}}^\top \text{diag}({\mathbf{x}_c}) \mathbf{x}_{u_c}\right)\\
\end{equation*}
Detailedly, $\mathbf{x}_{u_r} = \text{MLP}^U(\mathbf{E}^U_{u_r})$, $\mathbf{x}_{u_c} = \text{MLP}^U(\mathbf{E}^U_{u_c})$, and $\mathbf{x}_{c} = \text{MLP}^C(\mathbf{E}^C_c)$ where $\mathbf{E}^U$ and $\mathbf{E}^C$ are the user and content embeddings, $\mathbf{E}^U_{u_r}$ denotes the embedding for user $u_r$ and similar notations for the content side. $\mathbf{E}^U$ is learnable embedding while $\mathbf{E}^C$ is non-learnable and obtained from the text of posts via a sentence embedding model\footnote{Sentence-Transformer: https://sbert.net/}. To perform the matrix multiplications, the embeddings are all mapped to the same dimension through two multi-layer perceptron (MLP) for users and content respectively. The \textit{diag} operation transforms a vector into a square diagonal matrix. The final score is passed through sigmoid function to obtain the probability. The detailed model architecture is in the Appendix.

To train the model, the dataset is constructed based on the historical interactions. The format of each instance in the training dataset $\mathcal{D}$ is  $(u_r, u_c, c, p_c)$ where $p_c$ represents the probability of whether user $u_r$ will repost content $c$ created by user $u_c$. To build the instances, for each content $c$ posted by $u_c$, we collected the users who have reposted it as positive samples and the users who follow $u_c$ but do not repost $c$ as the negative instances. We assign $1$ for the positive instances and $0$ for the negative instances. There could be many negative instances if the creator has many followers, resulting in severe data imbalance and computational expenses. To avoid these issues, we randomly sample two users from the negative set.

We form this problem as a regression problem  and train the predictor parameterized by $\theta$ with the following Mean Square Error (MSE) loss:
\begin{equation}
    \mathcal{L} = \mathbb{E}_{(u_r, u_c, c, p_c) \in \mathcal{D}} (\mathcal{P}_\theta(u_r, u_c, c) - p_c)^2.
    \nonumber
\end{equation}

\subsection{Content Influence Estimation}

Given social network structure $G$ and probabilities 
derived from the pairwise influence estimator $\mathcal{P}_\theta(u_r, u_c, c)$, we estimate the influence of a user $u$ posting content $c$ (denoted as $I_u^c$) using the Independent Cascade Model~\cite{kempe2003maximizing}. In this work, we adopt the following relationship as the underlying network structure. 

The influence estimation is achieved through Monte Carlo simulations, where the process is repeated multiple times to obtain the average influence spread. In each simulation, user $u$ is the initial active node. The activation then propagates to neighbors based on the pre-determined probabilities from the pairwise influence estimator. The activations continue until no further activation happens. The number of activated nodes in a single simulation is represented by $N_i$. The influence $I_u^c$ of user $u$ posting content $c$ is calculated as the average number of activated nodes over $R$ simulations:
\begin{equation}
    I_u^c = \frac{\sum_{i=1}^{R} N_i}{R}.
    \label{eq.inf}
\end{equation}

\subsection{Problem Statement}

\begin{figure*}
    \centering

    \includegraphics[width=1\linewidth]{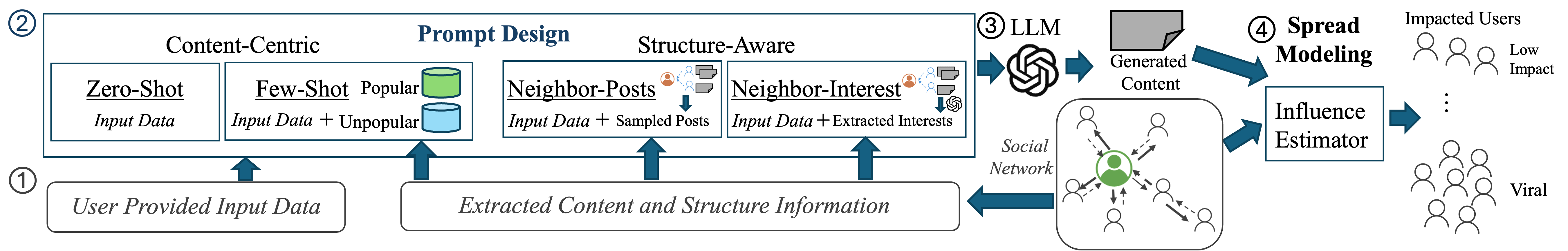}
    \vskip -1.5ex

    \caption{The overall framework consists of the following steps: (1) input data are collected based on prompt requirement; (2) prompts are generated using \textit{content-centric} and \textit{structure-aware} strategies; (3) the prompts are fed into LLMs to generate posts; and (4) the influence of the generated posts is evaluated through spread modeling.}
    
    \label{fig:framework}
    \vskip -3ex
\end{figure*}

In this paper, our focus is on generating content with substantial influence. Formally, a user $u$ can pre-define certain elements of the content they wish to generate, which are represented as $\mathcal{C}_u$. Examples of $\mathcal{C}_u$ include keywords, the initial part of a sentence, or even the entire sentence.\footnote{While our primary focus is on revising the original content to be more influential/popular, it's important to note that other elements can also be accommodated by modifying the prompts and inputs used in training.} This user-defined components are utilized to form the prompts according to various prompting strategies $(i.e., \mathcal{P}(\mathcal{C}_u))$. The generated content after feeding the prompt to a LLM $\mathcal{M}$ is denoted as $\mathcal{M}(\mathcal{P}(\mathcal{C}_u))$. Our goal is to obtain the best prompting strategy that can generate content of a high influence, measured as $I_u^c$. Formally, the objective is defined as:
\begin{equation}
\mathcal{P}^* = \underset{\mathcal{P}}{\mathrm{argmax}}\ \mathbb{E}_{u} \left[ I_u^{\mathcal{M}(\mathcal{P}(\mathcal{C}_u))} \right]
\label{eq.obj}
\end{equation}
Fig.~\ref{fig:framework} shows our framework to solve this problem. 

\section{Content-Centric Strategies}
\label{sec.initial}
In this section, we explore the prevalent \textit{in-context learning (ICL)} techniques in content generation to assess their impact on the influence of the generated posts. ICL is an advanced prompting strategy, which could potentially boost the downstream tasks~\cite{dong2022survey}. We explore commonly used zero-shot and few-shot settings. 

For zero-shot ICL, we define the scenario in social media and formalize the task to generate posts of high influence (\textit{Prompt 1}). For few-shot ICL, in addition to the provided information in the zero-shot setting, we provide several posts of high/low influence as examples (\textit{Prompt 2}). The posts used in the prompt are sampled from the training dataset where we heuristically treat the top $20\%$ posts sorted by influence\footnote{The reposting behavior in the real world is a reliable indicator of the post's influence. Therefore, for training posts, we use the repost number to measure the influence.} 
as posts of high influence and the bottom $20\%$ as posts of low influence. Based on whether the sampled posts are personalized to the input post, there are two variants:
\begin{itemize}[leftmargin=*]
    \item \textit{\textbf{Prompt 2.1} [fixed to all posts]:} the sampled instances are randomly drawn from the entire list of popular/unpopular posts in the training dataset and fixed for different inputs.
    \item \textit{\textbf{Prompt 2.2} [personalized to the input post]:} the sampled instances vary for input posts where the selected instances are the top similar posts in the textual embedding space.
\end{itemize}


\begin{tcolorbox}[colback=blue!10!white, colframe=blue!80!black, title=Prompt 1: Content-Centric-Zero-Shot, 
boxsep=0.75mm, 
  left=0.75mm, 
  right=0.75mm, 
  top=0.75mm, 
  bottom=0.75mm 
  ]
  \small
\textbf{Instruction:} Imagine you have a piece of text that you want to share on social media, but you want to ensure it catches the maximum attention and engagement from your audience. Your task is to creatively revise the original text to make it more engaging and more likely to be shared widely. The revised version should retain the core message but be optimized to resonate with social media trends and audience preferences. Your goal is to transform the text into a revised one that can lead to a larger cascade of shares, likes, and comments. 

\textbf{Input Data:} Input text=\textcolor{blue}{\{\textit{input post}\}}.
\end{tcolorbox}

\begin{tcolorbox}[colback=blue!10!white, colframe=blue!80!black, title=Prompt 2: Content-Centric-Few-Shot,boxsep=0.75mm, 
  left=0.75mm, 
  right=0.75mm, 
  top=0.75mm, 
  bottom=0.75mm 
  ]
  \small
\textbf{Instruction:} Imagine ... Your goal is to ...

\textbf{Demonstration:} Following are the examples of popular posts: \textcolor{blue}{\{\textit{sampled popular posts}\}}. Following are the examples of unpopular posts: \textcolor{blue}{\{\textit{sampled unpopular posts}\}}.

\textbf{Input Data:} Input text=\textcolor{blue}{\{\textit{input post}\}}.
\end{tcolorbox}

\section{Structure-Aware Strategies}
\label{sec.booster}

Content-centric strategies focus on the textual content but ignore the structural information, which will result in a sub-optimal performance in content dissemination. Inspired by the significant role of network structures in information spread~\cite{li2023IMsurvey}, we propose structure-aware strategies in this section to inject neighborhood information into prompt for generation.

Specifically, we augment prompt from $\mathcal{P}(\mathcal{C}_u)$ in Eq.~\ref{eq.obj} to $\mathcal{P}(\mathcal{C}_u, \mathcal{N}_u^h)$ to include the neighborhood information $\mathcal{N}$ within the specified network where $h$ is the number of hops and $\mathcal{N}_u^h$ are the users who needs at least $h$ steps for a content to spread from the creator $u$.
For instance, in the following network, the neighbors are those who follow the creator directly or can be reached via paths in the network. These target recipients $\mathcal{N}_u^h$, once receiving the post, could facilitate the information spread by reposting the message. Given the limitations imposed by prompt length, although the entire set of the historical reposts from the audience contains a full knowledge of their observed interests, it is impractical to incorporate all these posts into prompt. To best describe the neighborhood briefly, one can extract the representative posts from the audience's interactions, or directly summarize the audience's interests. Different ways are explored below and their prompts are listed as Prompt 3 and Prompt 4.

\begin{tcolorbox}[colback=orange!10!white, colframe=orange!80!black, title=Prompt 3: Structure-Aware-Neighbor-Posts,
boxsep=0.75mm, 
  left=0.75mm, 
  right=0.75mm, 
  top=0.75mm, 
  bottom=0.75mm 
  ]
  \small
\textbf{Instruction:} Imagine ... Your goal is to ...

\textbf{Neighborhood Information} Your audience has interacted with the following posts: \textcolor{orange}{\{\textit{posts from neighborhood interactions}\}}. Based on their preferences, now transform the text for higher popularity.

\textbf{Input Data:} Input text=\textcolor{orange}{\{\textit{input post}\}}.
\end{tcolorbox}

\begin{tcolorbox}[colback=orange!10!white, colframe=orange!80!black, title=Prompt 4: Structure-Aware-Neighbor-Interest,
boxsep=0.75mm, 
  left=0.75mm, 
  right=0.75mm, 
  top=0.75mm, 
  bottom=0.75mm 
  ]
  \small
\textbf{Instruction:} Imagine ... Your goal is to ...

\textbf{Neighborhood Information} Your audience has the following interest: \textcolor{orange}{\{\textit{summarized interest from neighborhood interactions}\}}. Based on their preferences, now transform the text for higher popularity.

\textbf{Input Data:} Input text=\textcolor{orange}{\{\textit{input post}\}}.
\end{tcolorbox}

\subsection{Representative Posts}
In Prompt 3, we inject the posts from neighborhood historical interactions to represent the interests of the potential audiences. Based on how the posts are selected, we have the following variants:
\begin{itemize}[leftmargin=*]
    \item \textit{\textbf{Prompt 3.0} [randomly from all training posts]}: posts are randomly selected from the whole training dataset.
    \item \textit{\textbf{Prompt 3.1} [randomly from the historical interactions of neighborhood]}: the content that the neighbors has interacted with indicates the interest, the posts are randomly sampled from their historical reposts.
    \item \textit{\textbf{Prompt 3.2} [randomly from the historical interactions of influential neighborhood]}: since an influential user has many audiences, targeting this user could potentially lead to a large cascade. Based on this, posts are sampled from the history of most influential audience.
\end{itemize}

\begin{figure*}[t]
    \centering
    \includegraphics[width=0.98\linewidth]{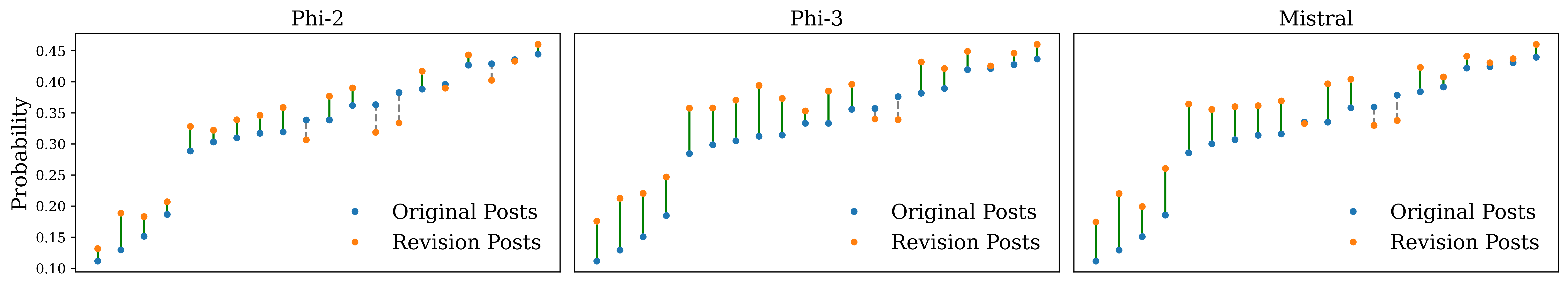}
    \caption{Probability comparison between the original post and the revised post.}
    \vspace{-0.5cm}
    \label{fig:prob_comparison}
\end{figure*}

\subsection{Summarized Interests}
Prompt 4 describes the neighborhood information by summarizing their interests. Specifically, a list of posts from the audience are first obtained and the interests from these posts are summarized using LLMs into a short sentence. The prompts for summarizing the interests are as follows. 

\begin{tcolorbox}[colback=olive!10!white, colframe=olive!80!black, title=Prompt: Interest Summarization, 
boxsep=0.75mm, 
  left=0.75mm, 
  right=0.75mm, 
  top=0.75mm, 
  bottom=0.75mm 
  ]
  \small
\textbf{Instruction:} Your goal is to summarize the interest of your audience based on their interactions and output the interests immediately, without any explanation or introduction. Your audience have interacted with the following posts: \textcolor{olive}{\{\textit{Neighborhood Posts}\}}. The response should be the interests and they are separated by `;'.
\end{tcolorbox}

Similarly, there are two variants based on the ways to obtain the neighborhood posts:
\begin{itemize}[leftmargin=*]
    \item \textit{\textbf{Prompt 4.1} [uniform sampling]}: firstly, a neighbor is randomly sampled from the neighborhood and then the post is randomly selected from this user's repost history. This process is repeated until reaching the defined number.
    \item \textit{\textbf{Prompt 4.2} [score-based sampling]}: the uniform method treats each audience equally, however, users will affect content spread differently. Intuitively, (1) if a user has more followers (i.e., higher degree), this user has a higher potential to increase the content's influence, (2) the importance of audience from 1-hop neighbors differs from that of 2-hop, 
    as the latter is influenced by the activation of the former based on the diffusion model. Based on these two factors, we use message passing in Graph Neural Networks (GNNs)~\cite{wu2020comprehensive}, which inherently utilizes the structure, as an aggregator to calculate the importance. More specifically, the user embedding is initialized as the binary encoding whose length equals user number and the element is set to 1 for user id otherwise 0. Then these embeddings are propagated based on a variant of non-parametric GCN to obtain the aggregated results with 
    $
        \mathbf{E}^{(l+1)} = \mathbf{D}^{\frac{1}{2}} \mathbf{A} \mathbf{D}^{\frac{1}{2}} \mathbf{E}^{(l)},
        \label{eq.gcn}
    $
    where $\mathbf{A}$ is the adjacency matrix and $\mathbf{D}$ is the diagonal degree matrix. The $i$-th position of the aggregated embedding is the importance of user $i$. Note that different than the traditional GCN where node with high degree is punished, we assign higher weights for high-degree node due to their importance in spreading the information.
    After having the importance, the audience is sampled based on the softmax probability and this process is repeated until reaching the pre-defined number.

\end{itemize}

\section{Experiments}

In this section, we aim to answer the following:
\begin{itemize}[leftmargin=*]
    \item \textbf{RQ1}: How well can the strategies amplify the influence of the generated post?
    \item \textbf{RQ2}: How effective is the influence estimator?
    \item \textbf{RQ3}: How does the neighborhood in the prompt affect the dissemination of the generated content?
    \item \textbf{RQ4}: How does user popularity relate to the content spread and the amplification effect through the strategies?
\end{itemize}

\subsection{Settings}
We conduct experiments on the publicly available Weibo
dataset\footnote{Weibo Dataset: http://aminer.org/Influencelocality} that is commonly used for influence study~\cite{qiu2018deepinf}. This dataset contains the raw text of the weibo posts and the following relationships between users as well as their post/repost behaviors. We use the repost interactions to train the pairwise influence estimator and use the following relationship as the underlying social network structure. For LLMs, we conduct experiments on the following models: Phi-2 and Phi-3 from Microsoft~\cite{abdin2024phi}, Mistral from Mistral AI~\cite{jiang2023mistral}. We also explore LLaMA 2 from Meta~\cite{touvron2023llama}, which exhibits strong ethical constraints and often refuses to perform the revision task. Consequently, we do not include it in this comparison. 
The details about the datasets, models, hyperparameters, and computing infrastructure are included in Appendix. Our code is available at 
\url{https://github.com/YuyingZhao/LLM-influence-amplifier}.

\subsection{RQ1: Effectiveness}
\label{sec.main_result}
To evaluate the effectiveness, we investigate from two perspectives. From a local probability perspective: targeting a single user, whether LLM can improve the probability of this user's repost behavior? Furthermore, given the current inference challenges, it is impractical to design personalized content for each viewer. A piece of content will be generated and disseminated to the audience rather than one individual. This inspires the global perspective. To evaluate how many users will be impacted globally, we measure the influence spread.

These two perspectives provide unique benefits. The local aspect (i.e., \textit{singular-user scenario}) isolates the challenge of aggregating interests from a user's network and focuses primarily on the individual repost probability. The global aspect (i.e., \textit{influence spread}) requires dedicated consideration of multiple simulations and reposting rounds as well as a broader range of users, thus presenting the more complex scenario.

\subsubsection{Singular-User Scenario}
The local perspective evaluates a single tuple $(u_r, u_c, c)$ to assess whether user $u_r$ will repost content $c$ posted by user $u_c$. 
Ideally, content tailored specifically to an individual user is hypothesized to be more influential.
To empirically test this hypothesis and verify the effectiveness of LLM in improving the influence on singular user, we employ LLM prompted with content $c$ and randomly selected reposts from user $u_r$ that reflect their content preferences according to Prompt 3. The effectiveness is assessed by analyzing 20 posts processed by different LLMs, as illustrated in Fig.~\ref{fig:prob_comparison}. For each post, we compute the average influence probability over 20 randomly chosen neighbors. For better clarity of the visualization, The arrangement of posts is organized by the repost probability prior to revision. Vertical lines connect the pre- and post-revision states of each post, where a solid green line indicates the revision increase in influence probability, and a grey dashed line indicates a decrease. The majority of revised posts demonstrate an increased probability, thus supporting the efficacy of the LLM. Specifically, the success rates for increasing influence probability after the revision are $70\%$ for Phi-2, $90\%$ for Phi-3, and $85\%$ for Mistral, indicating a substantial potential of LLMs to enhance content impact in single-user scenarios.

\subsubsection{Influence Spread: Content-Centric}

\begin{table}[t]
\centering
\small 
\caption{Content-Centric: influence spread.}
\vspace{-0.1cm}

\resizebox{0.45\textwidth}{!}{%
\begin{tabular}{|c|c|c|c|}
\hline
& \multicolumn{1}{c|}{Zero-Shot} & \multicolumn{2}{c|}{Few-Shot} \\ \hline
 & Prompt 1 & Prompt 2.1 & Prompt 2.2 \\ \hline
Phi-2 & 122.97 (+0.86\%) & 137.75 (+12.99\%) & 133.96 (+9.88\%) \\ \hline
Phi-3 & 140.43 (+15.18\%) & 145.24 (+19.13\%) & 144.83 (+18.79\%) \\ \hline
Mistral & 139.18 (+14.16\%) & 141.68 (+16.2\%) & 139.31 (+14.27\%) \\ \hline
\textbf{Average} & \textbf{+10.73\%} & \textbf{+16.77\%} & \textbf{+14.31\%} \\ \hline
\end{tabular}%

}
\vspace{-0.3cm}
\label{tab.content_centric}
\end{table}

\begin{table*}[h]
\centering
\footnotesize
\caption{Structure-Aware: average influence spread across different prompts.}
\vspace{-0.1cm}
\resizebox{0.8\textwidth}{!}{%
\begin{tabular}{|c|c|c|c|c|c|}

\hline
& \multicolumn{3}{c|}{Sampled Posts} & \multicolumn{2}{c|}{Summarized Interests} \\ \hline
& Prompt 3.0 & Prompt 3.1 & Prompt 3.2 & Prompt 4.1 & Prompt 4.2 \\ \hline
Phi-2 & 135.79 (+11.37\%) & 136.62 (+12.06\%) & 131.88 (+8.17\%) & 131.44 (+7.81\%) & 129.0 (+5.81\%) \\ \hline
Phi-3 & 148.48 (+21.79\%) & 146.69 (+20.32\%) & 144.69 (+18.67\%) & 144.93 (+18.87\%) & 143.44 (+17.65\%) \\ \hline
Mistral & 145.42 (+19.27\%) & 141.88 (+16.37\%) & 140.68 (+15.39\%) & 143.49 (+17.69\%) & 139.16 (+14.14\%) \\ \hline
\textbf{Average} & \textbf{+17.48\%} & \textbf{16.25\%} & \textbf{+14.08\%} & \textbf{+14.79\%} & \textbf{+12.53\%} \\ 
\hline

\end{tabular}%
 }
\vspace{-0.1cm}
\label{tab.structure_aware}
\end{table*}

The influence spread is calculated according to Eq.~(\ref{eq.inf}). The average spread of the original posts is $121.92$. The results for different prompts and LLMs are reported in Table~\ref{tab.content_centric}, which also includes the relative gain compared with the original posts. The table provides several key insights:

\noindent \textbf{LLM Comparison}: All combinations of the LLMs and prompts result in an enhanced influence spread over the original posts, confirming the broad efficacy of LLMs in amplifying content impact. Notably, within the same series, Phi-3 model, which possesses more parameters, outperforms its smaller counterpart, Phi-2. Conversely, despite having a larger parameter set, Mistral slightly under-performs Phi-3, potentially due to specific aspects of its architectural design or training dataset. Among the prompts, Phi-3 consistently shows superior performance. This observation also aligns with singular-user scenario where Phi-3 has the highest ratio of improving the probability.
    
\noindent \textbf{Prompts Comparison}: All prompts consistently improve the influence spread across LLMs, validating the effectiveness of the LLM-aided content generation for higher popularity/influence. Prompts utilizing multiple examples (few-shot) show greater improvements over their counterparts without examples (15.54$\%$ vs. 10.73$\%$), underscoring the effectiveness of including examples. Moreover, prompts that sample globally from popular/unpopular posts (Prompt 2.1) prove to be more effective than those sampling locally based on the input content (Prompt 2.2), highlighting the value of a broader sampling strategy.

\subsubsection{Influence Spread: Structure-Aware}
Similarly, we report the average influence spread for the structure-aware prompts in Table~\ref{tab.structure_aware}. Note that these prompts are based on Prompt 1 (zero-shot) and they can be used simultaneously with the few-shot strategies. Compared with the Prompt 1 with an average of 10.73$\%$ improvement, the prompts with the structural information consistently have a better improvement. This shows that incorporating the neighborhood information is beneficial in boosting the influence. 
Additionally, directly incorporating the sampled posts (Prompt $3.*$) is more effective than the summarized interests (Prompt $4.*$). Targeting the most influential audience does not guarantee a better improvement as evidenced by Prompt 3.2 and Prompt 4.2.

\subsection{RQ2: Effectiveness of Influence Estimator}
\label{sec.influence_estimator}

\begin{figure}[t]
    \centering
    \includegraphics[width=0.48\textwidth]
    {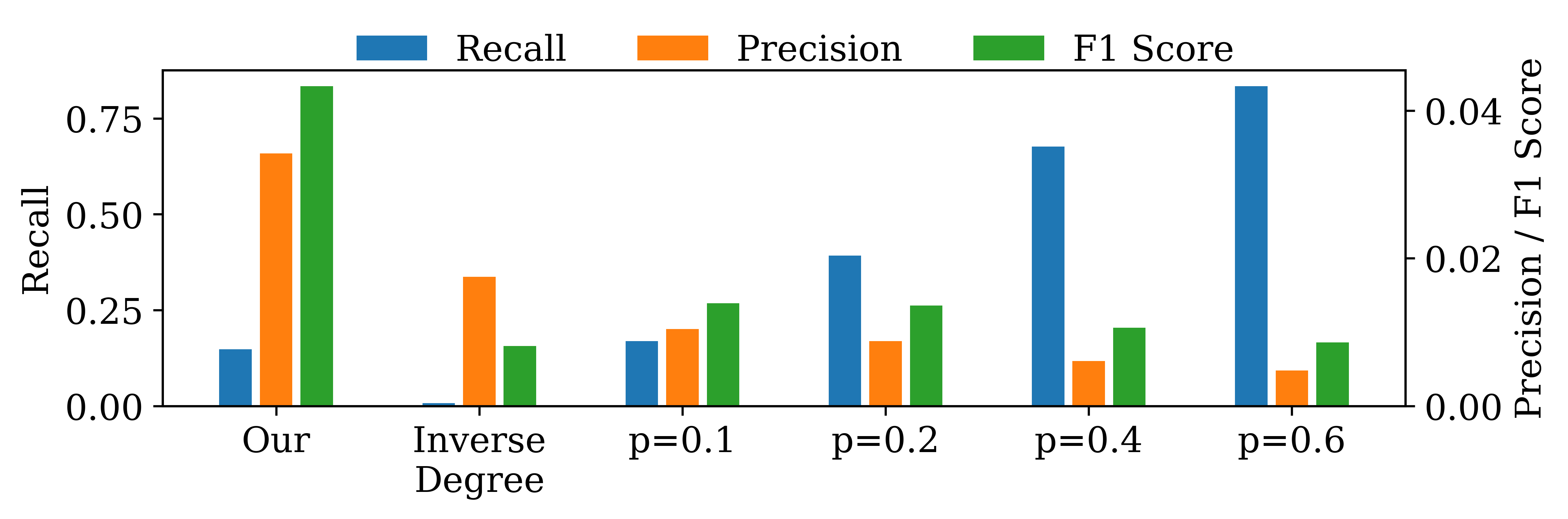}
    \vspace{-0.75cm}
    \caption{Influence evaluation performance comparison.}
    \vspace{-0.25cm}
    \label{fig.coefficient}
\end{figure}

To verify the effectiveness of the influence estimator, we compare its performance with different variants based on the same independent cascade model. These compared methods differ in the probability assignment to the edges in the network. Our content-aware evaluator learns the probability based on the historical interactions. We compare other probability assignments including (1) uniformly assigning a fixed probability to all edges, with explored values ranging from 0.1 to 0.6, and (2) employing the inverse of node degree as the probability, a strategy to mimic the user's behavior where individuals following numerous others are less likely to further propagate a message.

One inherent challenge in measuring the performance is the absence of concrete `ground truth' data regarding the influence of generated content, as these contents have not been disseminated in real-world settings. However, the collected posts have actual repost behavior in practice, providing a reliable indicator of the influence. Therefore, for the collected posts, we treat the users who reposted the post as the positive ground truth and those who are within the following network without repost behavior as the negative ground truth. Based on the ground truth, we calculate the Recall, Precision, and F1 for the simulations and get the average scores. Fig.~\ref{fig.coefficient} presents the results. From the figure, we notice a clear trend for the methods with fixed probabilities. As probability increases, the recall will improve and the precision will decrease. This trend is expected, as higher probabilities typically yield a greater number of positive instances. In comparison, methods employing inverse degree adjustments exhibit higher precision but lower recall. This is attributed to the reduced probabilities associated with high-degree users. In contrast, our method has the best F1 score when compared with the others. This underscores the significant role of content in influencing spread estimation. Our content-aware approach more reliably estimates influence probabilities, thereby enhancing the accuracy of influence estimation.

\subsection{RQ3: Impact of Neighborhood}
\label{sec.impact_hop}

\begin{figure}[t]
    \centering
    \vspace{-0.2cm}
    \includegraphics[width=0.4\textwidth]{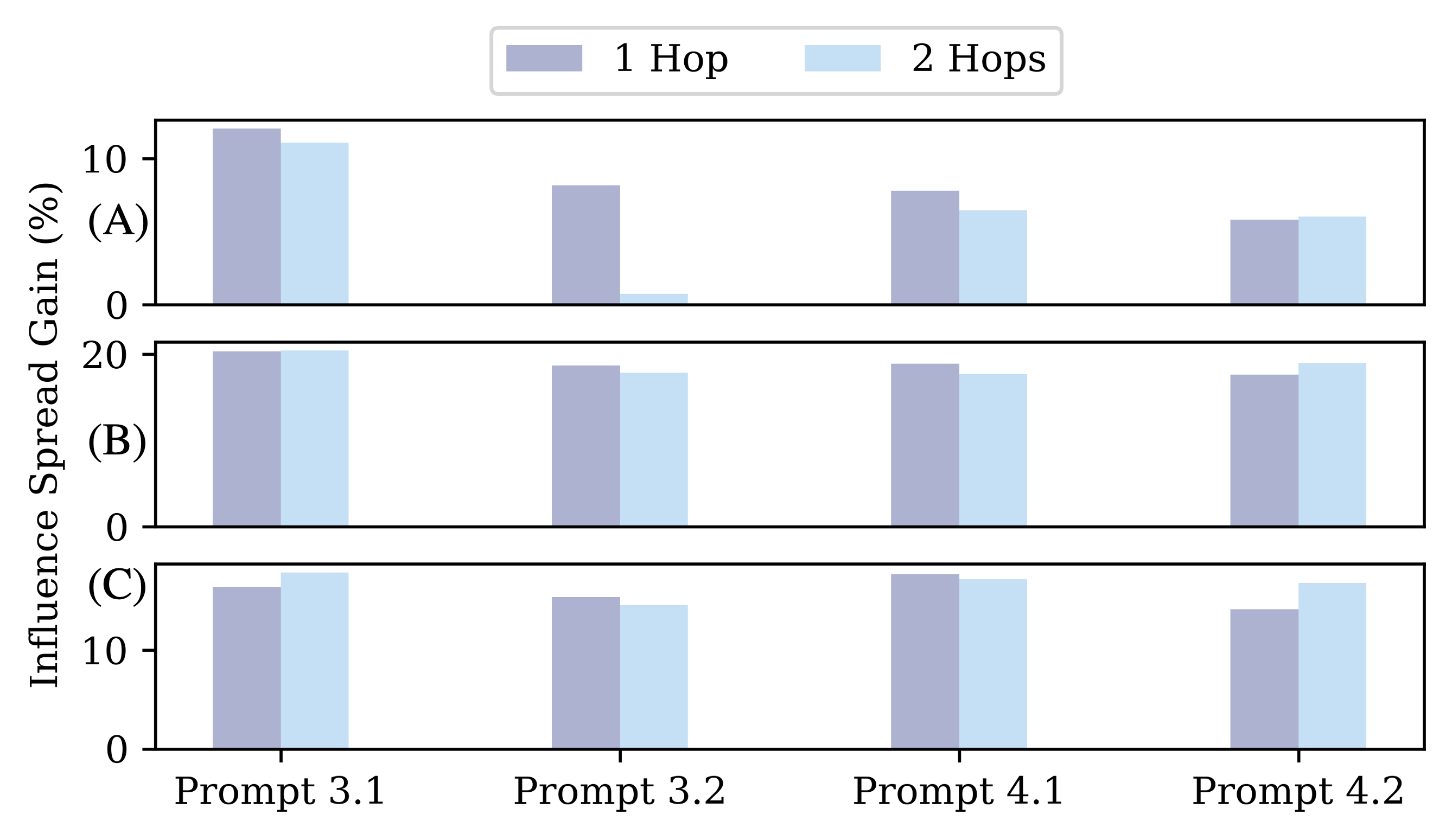}
    \caption{Performance gain comparison for different hops: (A) Phi-2, (B) Phi-3, (C) Mistral.}
    \vspace{-0.4cm}
    \label{fig.gain}
\end{figure}

The number of hops determines the creator neighborhood and will affect the prompts that use neighborhood information, including posts-based prompts 3.1 and 3.2, and interest-based prompts 4.1 and 4.2. The results are shown in Fig.~\ref{fig.gain}. For Prompt 3.1 and 4.1 which are both based on randomly sampled posts, they are not affected significantly. For Prompt 3.2 and 4.2 that favor influential audience, there is a varied trend as the number of hops increases. For post-based prompt, the performance drops for all LLMs including a significant drop for Phi-2. The interest-based one shows a consistent improvement. Comparing prompt 4.1 and 4.2, influential-based sampling is better than random-based sampling for multiple hops.

\subsection{RQ4: Analysis Across User Groups}
\label{sec.fair_study}
\begin{figure}
    \centering
    \includegraphics[width=0.9\linewidth]{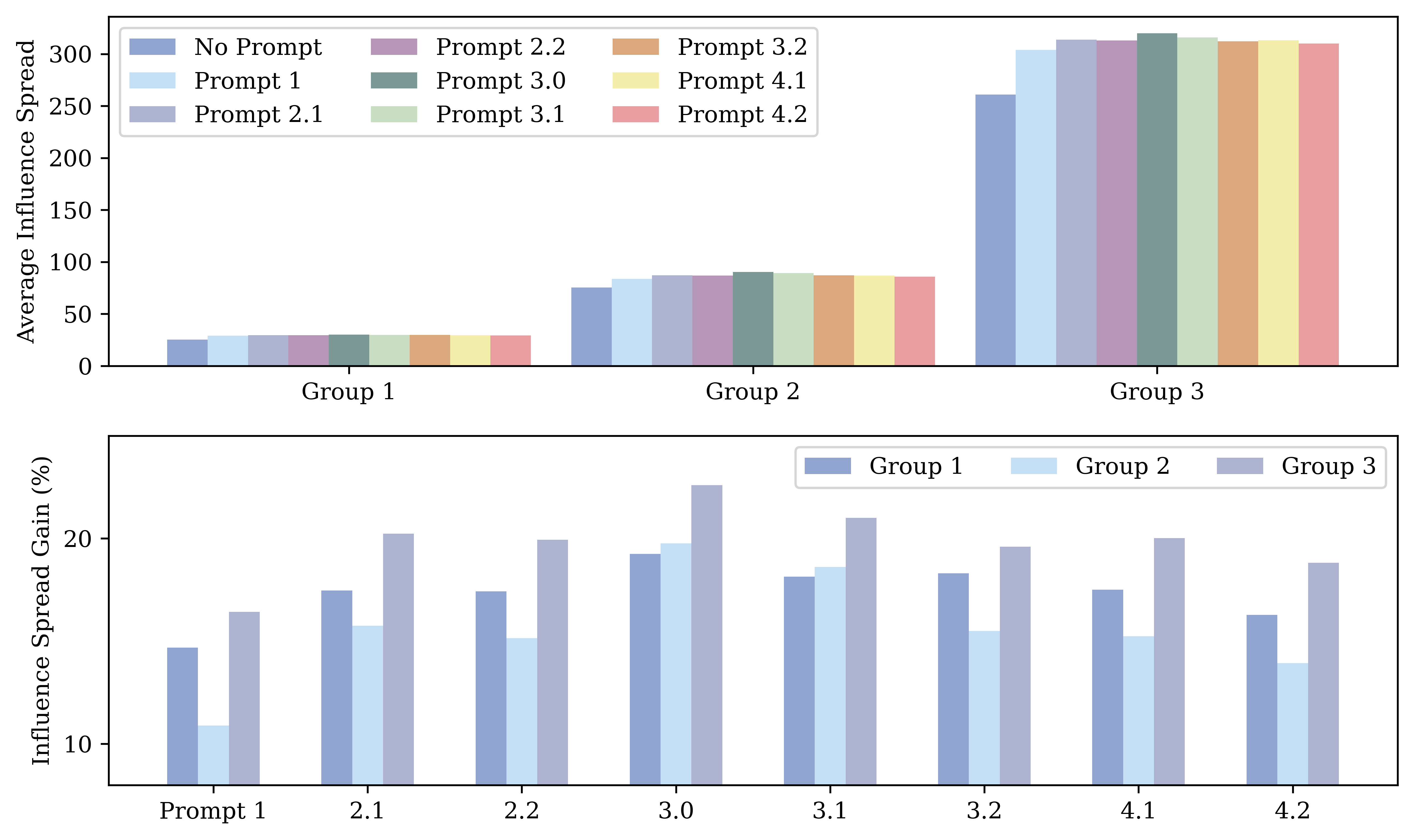}
     \vspace{-0.1cm}
    \caption{Group-level influence spread and gain (Phi-3).}
    \vspace{-0.5cm}
    \label{fig:group}
\end{figure}

We group users into three even groups based on follower numbers. Group 1 has the lowest degree and Group 3 has the highest. Results are shown in Fig.~\ref{fig:group}. The influence spread per group shows that the influence spread is highly related to the user degree where the higher degree users tends to have larger influence, aligning with real-world observation. We also plot the relative gain compared with original post. Generally the users with low and high degrees have a larger improvement. For low degree users it is possible their initial spread is low and the ratio for improvement becomes high. For high degree users, despite their initial large spread, the improvements are the highest. This suggests that cautions need to be paid in practice for such revising tool which might enlarge the difference between popular and unpopular creators. The results for two other LLMs are in the Appendix.

\section{Conclusion}
\label{sec.conclusion}

In this work, we investigate the intriguing question: \textit{Can LLMs generate personalized influential content to amplify a user’s presence on social media?} We propose a content-aware influence estimator to evaluate the influence of generated content and design diverse content-centric and structure-aware prompts. Through empirical experiments, we validate that LLMs can improve the content influence and draw insights on which prompts are more effective. In the future, we aim to optimize LLMs to generate influential content from a model perspective. While fixing the best prompt strategy, how to find the optimal model poses new challenges.


\newpage
\section*{Limitations and Discussions}
Our study aims to help benign users amplify their social presence. While these methods are promising and can potentially help many users to reach their goal, they also present risks, as they could be a double-edged sword. We acknowledge the potential for misuse by malicious users to spread misinformation. In practice, similar strategies might already be in use by these users to spread misinformation, even without our research. One limitation of this work is the absence of direct strategies to counter such misuse. However, our work can serve as an essential first step by simulating how adversaries might devise and execute such tactics. Existing works on misinformation prevention mainly focuses on detection. However, preventing misinformation spread is as crucial as designing the detection. In the age of LLMs, the quality of the generated misinformation has been improved by leveraging the open-world knowledge encoded in LLMs, which inherently increases the challenges of detecting misinformation. Furthermore, the cost of generating misinformation is largely reduced and malicious users now can produce and spread higher-quality misinformation at an unprecedented rate. Therefore, the content evading detection, even if of a small percentage, will result in a substantial volume of misleading posts online. Such spread of misinformation poses a high threat to society. By understanding and replicating potential adversarial behavior, this research lays a strong foundation for future initiatives focused on developing effective countermeasures against misinformation dissemination.

In this study, we focus on explicit following relationships as the foundation of the social network. However, on modern social media platforms, information can also be disseminated through recommendations, where content is suggested to potential recipients by the platform. Therefore, in practice, it is possible to extract a recommendation network and combine it with the following network to analyze information propagation in a more comprehensive and realistic manner. While our framework can be extended to the recommendation or merged networks, the findings presented here are restricted to the following network.
\vspace{8ex}

\bibliography{custom}

\appendix
\newpage
\section{Appendix}

\subsection{Predictor Architecture}
\begin{figure}[h]
    \centering
    \includegraphics[width=0.99\linewidth]{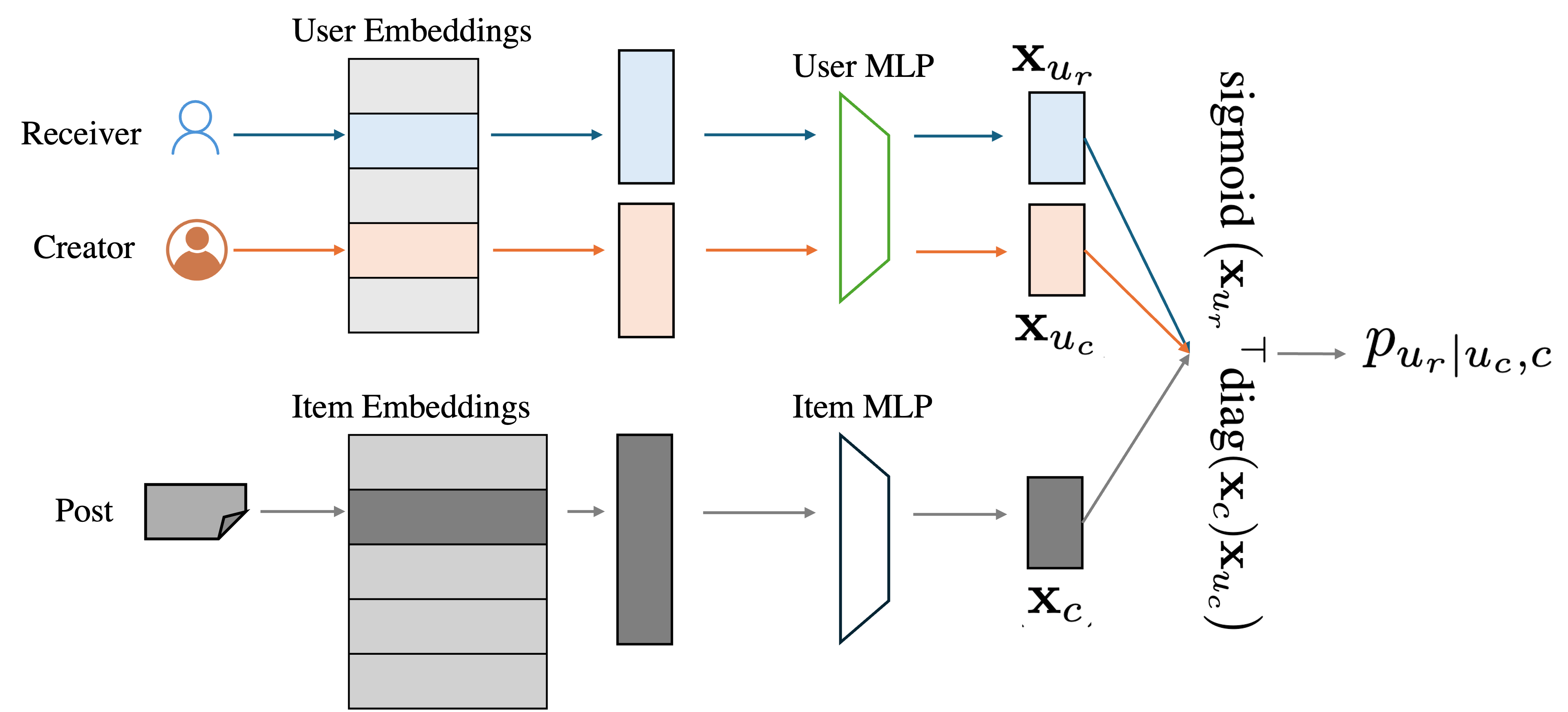}
    \caption{The pipeline of pairwise influence predictor.}
    \label{fig:predictor}
\end{figure}

According to the equation of pairwise influence estimator:
\begin{equation*}
    \mathcal{P}_\theta(u_r, u_c, c) = \text{sigmoid}\left({\mathbf{x}_{u_r}}^\top \text{diag}({\mathbf{x}_c}) \mathbf{x}_{u_c}\right),\\
\end{equation*}
the pipeline is illustrated in Fig.~\ref{fig:predictor}. The user embeddings are randomly initialized embeddings which are learnable. Item embeddings are pre-computed by feeding the posts into sentence transformer (specifically, we used pre-trained sentence model \textit{sentence-transformers/distiluse-base-multilingual-cased-v2}\footnote{The url of the specific sentence transformer model that is used in the predictor: https://huggingface.co/sentence-transformers/distiluse-base-multilingual-cased-v2}). They are fixed throughout the training. There are two MLPs to transform the embeddings into the same dimension for users and items separately. Both of them have one hidden layer. The dimensions in user MLP are $[32, 64, 32]$ for input layer, hidden layer, and output layer, and $[512, 128, 32]$ for item MLP. A dropout of $0.5$ has been applied to the layers to prevent over-fitting. The batch size of training is set to 1024 and the learning rate is $1e^{-4}$.

\subsection{Dataset}
The original weibo dataset is downloaded from http://aminer.org/Influencelocality. The original data contains many files, we use the following 
in our study since they contain all the necessary information we need to conduct the investigation.
\begin{itemize}[leftmargin=*]
    \item \textit{root\_content.txt}: it contains the raw textual content of the posts;
    \item \textit{weibo\_network.txt}: it contains the following relationships between users;
    \item \textit{total.txt}: it records the post and repost beheviors (e.g., who created which post and who reposted which post).
\end{itemize}
To mitigate the sparsity issues in the original dataset, we perform the following preprocessing steps. We use the authors of the top 20 mostly reposted content as the initial seed $\mathcal{U}_\text{seed}$ and extract the relevant users who has reposted $\mathcal{U}_\text{seed}$' posts and the users that $\mathcal{U}_\text{seed}$ have reposted. These users become the new seed. We repeat this process two times to obtain the relatively dense users where they have probabily interacted with the others in the network. Additionally, only users having at least one post behavior and one repost behavior have participated in the above process. The network structure is extracted based on the obtained user subset where edges $(u, v)$ are kept when $u$ and $v$ both fall in the user subset. We further conduct the pruning from the content perspective. When investigating the spread of posts, we focus on posts that have at least been reposted by 5 users within the users' following network. The motivation behind this is that we focus on the following relationship in this study while the methodology can be extended to other networks (e.g., implicit recommendation network learned from historical interactions). The post subset is splitted into train/val/test based on $60\%/20\%/20\%$ proportion. After the prepossessing steps, there are 13863 users and 6112 posts for the influence investigation.

\subsection{Hyperparameters}
For predictor training, we set learning rate as $1e^{-4}$, batch size as 1024, the dimension of user embedding as 32. For prompt 2.1 and 2.2, we sample two popular posts and two unpopular posts. For the post-based prompts (3.1 and 3.2), the inserted post number is set to 3. For the interest-based prompts (4.1 and 4.2), there are 10 posts initially sampled from the neighborhood to do the interest extraction. We experienced out of memory issue when sampling more posts in the current server. The number of simulation rounds during influence evaluation is set to 20.

\begin{figure*}[ht]
    \centering
    \begin{subfigure}[b]{0.48\textwidth}
        \centering
        \includegraphics[width=\textwidth]{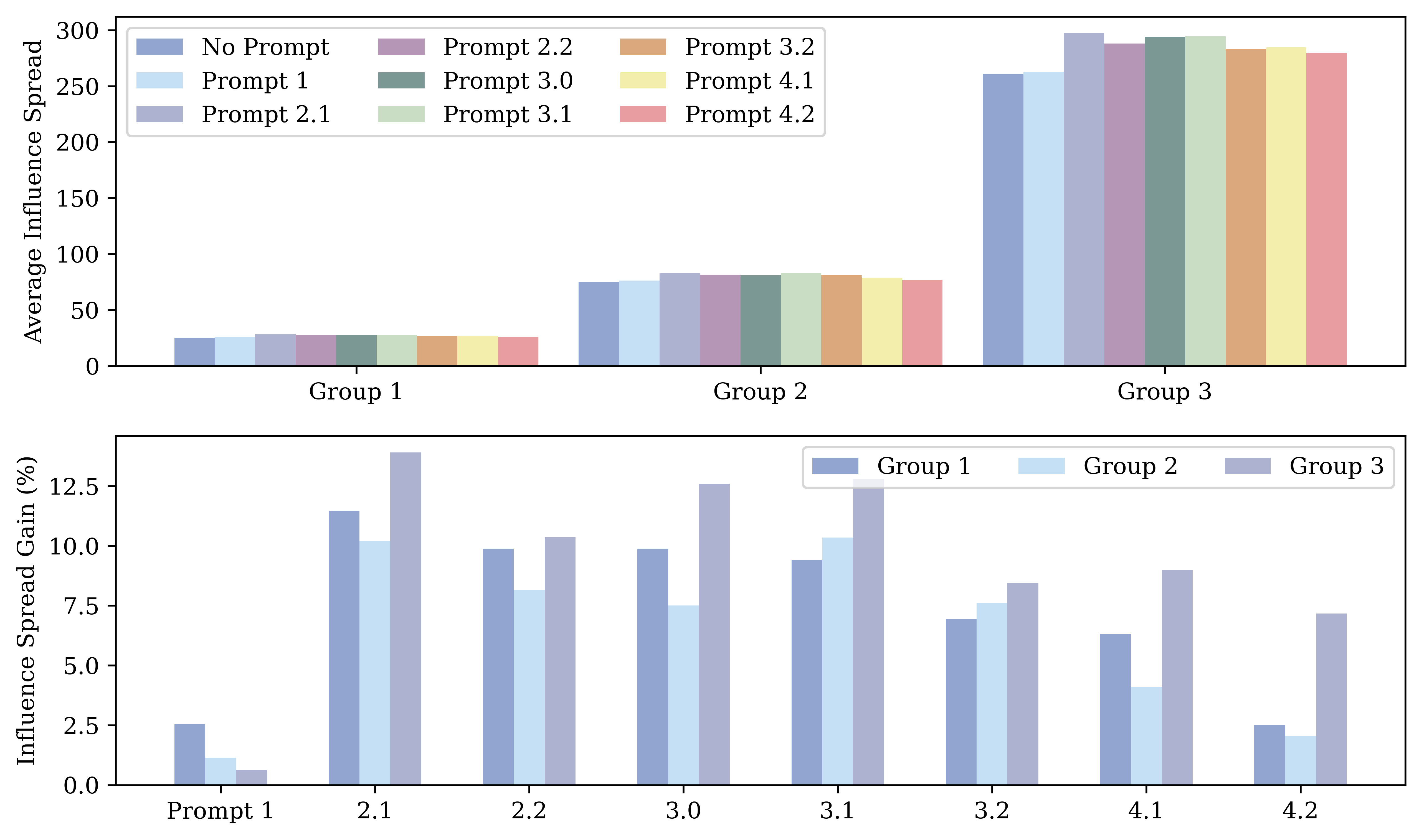}
        \caption{Phi-2}
        \label{fig:sub1}
    \end{subfigure}
    \hfill
    \begin{subfigure}[b]{0.48\textwidth}
        \centering
        \includegraphics[width=\textwidth]{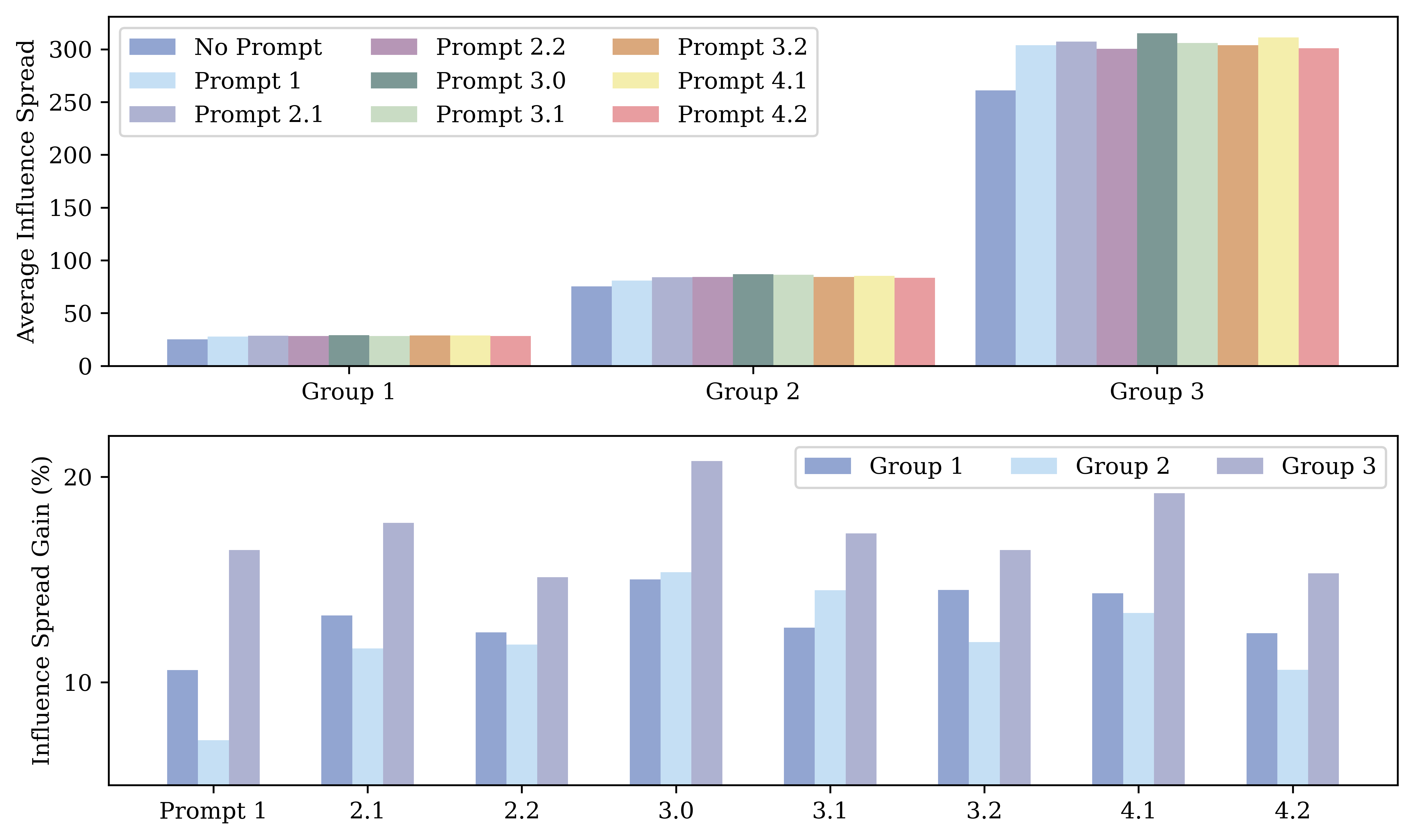}
        \caption{Mistral}
        \label{fig:sub2}
    \end{subfigure}
    \caption{Group-level average influence spread and relative gain on Phi-2 and Mistral models.}
    \label{fig:group_results_phi2_mistral}
\end{figure*}

\subsection{Computing Infrastructure}
The experiments in this work were conducted using an NVIDIA GeForce RTX 3090 GPU with 24GB memory. The operating system was Ubuntu 22.04.4 LTS. In terms of the software, our code for LLM part is mainly based on the popular framework called LitGPT\footnote{LitGPT: https://github.com/Lightning-AI/litgpt} that implemented 20+ high-performance LLMs with recipes to pretrain, finetune and deploy at scale. An easy installation of the packages can be achieved by \textit{pip install `litgpt[all]'}. Since we use sentence transformer to compute the post embedding during pairwise influence predictor training, also need to download the package using \textit{pip install -U sentence-transformers}. Detailed descriptions of the software environment and code are provided in the readme in the zipped code file.

\subsection{LLMs}
In our current work, we compare three models, Phi-2\footnote{Phi-2: https://huggingface.co/microsoft/phi-2}, Phi-3\footnote{Phi-3: https://huggingface.co/microsoft/Phi-3-mini-4k-instruct} from Microsoft, and Mistral\footnote{Mistral: https://huggingface.co/mistralai/Mistral-7B-Instruct-v0.3} from Mistral AI. The number of parameters in three models are 2.7B, 3.8B and 7B. In the future, we plan to explore more LLMs.

\subsection{Group-Level Results}

The average influence spread and the relative improvement in percent are shown in Fig.~\ref{fig:group_results_phi2_mistral} for the other two models: Phi-2 and Mistral. Firstly, they show the same trend for the average influence spread with the observation in main content for Phi-3, where users with high degree have a higher influence spread. Additionally, except Prompt 1 in Phi-2, the other prompts across models all have the best improvement for group 3 with a high degree.
\end{document}